\documentclass[runningheads]{llncs}

\usepackage{cite}
\usepackage{graphicx}
\usepackage{amsmath}
\usepackage{booktabs}

\usepackage{svg}
\usepackage{subfigure}
\usepackage{amsfonts}
\usepackage{amssymb}
\usepackage{booktabs}
\usepackage{marvosym}
\usepackage{hyperref}
\usepackage{floatrow}
\newfloatcommand{capbtabbox}{table}[][\FBwidth]
\newfloatcommand{figurebox}{figure}[][\FBwidth]

\begin{document}

\title{LSSANet: A Long Short Slice-Aware Network for Pulmonary Nodule Detection}
\titlerunning{LSSANet: A Long Short Slice-Aware Network}

\author{Rui Xu\inst{1,2} \and
Yong Luo\inst{1,2}\thanks{Yong Luo and Bo Du are the corresponding authors.} \and
Bo Du\inst{1,2\star} \and
Kaiming Kuang \inst{4} \and
Jiancheng Yang \inst{3,4,5}}

\authorrunning{R. Xu et al.}

\institute{National Engineering Research Center for Multimedia Software, School of Computer Science, Institute of Artifical Intelligence, and Hubei Key Laboratory of Multimedia and Network Communication Engineering, Wuhan University, Wuhan, China \\ \and Hubei Luojia Laboratory, Wuhan, China \\
\email{\{luoyong, dubo\}@whu.edu.cn} \\
\and
Shanghai Jiao Tong University, Shanghai, China \\ \and
Dianei Technology, Shanghai, China\\ \and
EPFL, Lausanne, Switzerland}

\maketitle 

\begin{abstract}
Convolutional neural networks (CNNs) have been demonstrated to be highly effective in the field of pulmonary nodule detection. However, existing CNN based pulmonary nodule detection methods lack the ability to capture long-range dependencies, which is vital for global information extraction. In computer vision tasks, non-local operations have been widely utilized, but the computational cost could be very high for 3D computed tomography (CT) images. To address this issue, we propose a long short slice-aware network (LSSANet) for the detection of pulmonary nodules. In particular, we develop a new non-local mechanism termed long short slice grouping (LSSG), which splits the compact non-local embeddings into a short-distance slice grouped one and a long-distance slice grouped counterpart. This not only reduces the computational burden, but also keeps long-range dependencies among any elements across slices and in the whole feature map. The proposed LSSG is easy-to-use and can be plugged into many pulmonary nodule detection networks. To verify the performance of LSSANet, we compare with several recently proposed and competitive detection approaches based on 2D/3D CNN. Promising evaluation results on the large-scale PN9 dataset demonstrate the effectiveness of our method. Code is at \url{https://github.com/Ruixxxx/LSSANet}.

\end{abstract}

\section{Introduction}
Lung cancer is the leading cause of cancer-related death worldwide. Prompt diagnosis and timely treatments of the pulmonary nodules are critical solutions for lung cancer, which can significantly improve the prospects of survival. In order to realize the pulmonary nodules' early diagnosis, thoracic computed tomography (CT) is widely adopted and demonstrated to be an effective tool. Nonetheless, manually identifying nodules in CT images is a labor-intensive task, since interpreting CT data requires doctors to analyze hundreds of slices every time. 
Hence, it is desirable to introduce machine learning to automatically and accurately identify and diagnose the pulmonary nodule.

In recent years, with the prosperity of deep learning, convolutional neural networks (CNNs) have been introduced to assist doctors in the field of pulmonary nodule detection \cite{DBLP:journals/tmi/SetioCLGJRWNSG16,DBLP:conf/miccai/DingLHW17,DBLP:journals/tbe/DouCYQH17,deeplung,DBLP:journals/nn/KimYCS19,leaky_noisy-or,nodulenet,i3dr-net,deepseed,DBLP:journals/tmi/OzdemirRB20,DBLP:conf/miccai/SongCLHLHCYSZW20,sanet}. 
Owing to the 3D nature of CT images, for approaches based on 2D CNNs \cite{DBLP:journals/tmi/SetioCLGJRWNSG16,DBLP:conf/miccai/DingLHW17}, post-processing is often required to integrate the 2D regions detected into 3D, but this may affect the efficiency and also accuracy of the pulmonary nodule detection. Therefore, 3D CNN based methods \cite{DBLP:journals/tbe/DouCYQH17,deeplung,DBLP:journals/nn/KimYCS19,leaky_noisy-or,nodulenet,i3dr-net,deepseed,DBLP:journals/tmi/OzdemirRB20,DBLP:conf/miccai/SongCLHLHCYSZW20,sanet} are dominant in this field, where most state-of-the-art frameworks consist of three components: U-Net-like backbone \cite{unet15}, 3D region proposal network, and the false positive reduction module. However, the aforementioned methods lack the ability to capture long-range dependencies, which is vital for global information extraction in vision tasks. Compact non-local operation \cite{non-local18,compact_non-local18} is able to alleviate this issue and improve the discrimination ability of fine-grained objects. However, when we directly apply such operations to pulmonary nodule detection, the computational or memory cost may be tremendous resulting from the 3D nature of CT images.

A possible solution to reduce the computational cost is to employ the idea of grouping, which is studied by many existing works, such as Xception \cite{Xception17}, MobileNet \cite{mobilenets17}, ResNeXt \cite{ResNeXt17}, Group normalization \cite{GN18}, and CrossFormer \cite{crossformer21}. Since the computational complexity is usually nonlinear w.r.t. the input size, dividing the input into groups has been demonstrated to be effective in reducing the computational cost. Considering the fact that one nodule usually exists in several adjacent slices, we follow the way of grouping in \cite{sanet}, which groups the neighboring embeddings by depth. Yet, the adjacent depth grouping operations will lose the correlations among elements across long-distance slices, and hence may undermine the elimination of continuous pipe-like structures and the localization of pulmonary nodules.

Therefore, we propose a substitute of the vanilla compact non-local mechanism termed long short slice grouping (LSSG). In particular, we split the compact non-local operation into short-distance slice grouping (SSG) and long-distance slice grouping (LSG). SSG captures the long-range dependencies among any elements of adjacent slices, while LSG takes charge of the dependencies among any elements of slices far away from each other. More importantly, the proposed LSSG can further improve the performance by focusing on the small-scale features of nodules existing in consecutive slices, and realizing global feature extraction from the sparse slices.

Based on the LSSG introduced above, we construct a new end-to-end framework termed LSSANet, for the detection of pulmonary nodules. Following \cite{sanet}, 3D ResNet50 \cite{ResNet16} followed by two deconvolutional layers is adopted as a U-shaped encoder-decoder backbone \cite{unet15}. Our proposed LSSG is integrated into the encoder to improve the feature extraction. Then, a 3D region proposal network is employed on the output of the backbone to generate pulmonary nodule candidates. Finally, multi-scale regions of interest are fed into one false positive reduction network to further decrease false positives.

To summarize, the main contributions of this paper are:
\begin{itemize}
    \item We develop a long short slice-aware network (LSSANet) aiming at pulmonary nodule detection, where a novel long short slice grouping (LSSG) module is designed to simultaneously reduce the computational cost and capture the long-distance dependencies.
    \item Multiple variants of the designed LSSG by placing different numbers of LSG and SSG operations in different orders in the encoder are compared.
\end{itemize}
We conduct experiments on the large-scale PN9 \cite{sanet} dataset, and the results demonstrate that our proposed LSSANet can significantly outperform the state-of-the-art 2D CNN based and 3D CNN based pulmonary nodule detection approaches.

\begin{figure}[t]
\centering
\includegraphics[width=4.4in]{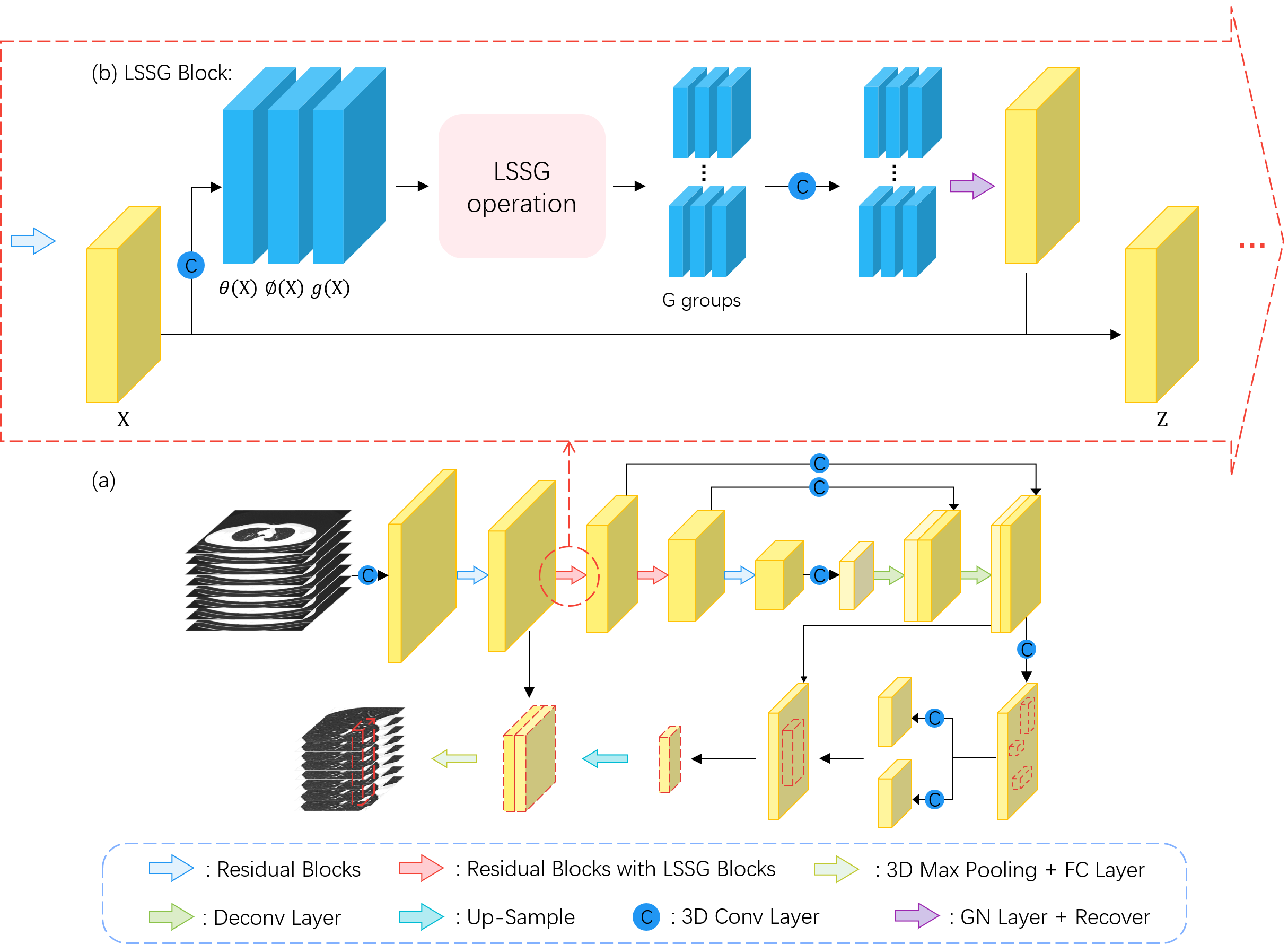}
\caption{(a) An illustration of our long short slice-aware network (LSSANet), where a designed long short slice grouping (LSSG) block is integrated into the encoder of U-shaped backbone. 
(b) The LSSG block. Embeddings resulting from three $1\times1\times1$ convolutional layers are grouped into $G$ slice groups using our proposed LSSG operation. Then 3D convolution and GN are applied for different groups. After depth recovery and integrating with the original representation, we obtain an improved representation that explores the long-range dependencies among the elements across long/short-distance slices.}
\label{LSSANet}
\end{figure}

\section{Methods}
Inspired by the tremendous success of non-local mechanism \cite{non-local18} for modeling long-range dependencies in computer vision, and to take full advantage of the CT images' 3D nature, we develop a long short slice-aware network (LSSANet). The overall architecture of LSSANet is illustrated in Fig. \ref{LSSANet}, which comprises an encoder-decoder backbone, a 3D region proposal component, and a multi-scale false positive reduction component. Particularly, we design a long short slice-grouping (LSSG) module to explore the long-range dependencies among any elements of one long/short-distance sliced group in the feature map and is integrated into the encoder.

\subsection{Long Short Slice Grouping}
\begin{figure}
\centering
\includegraphics[width=4.5in]{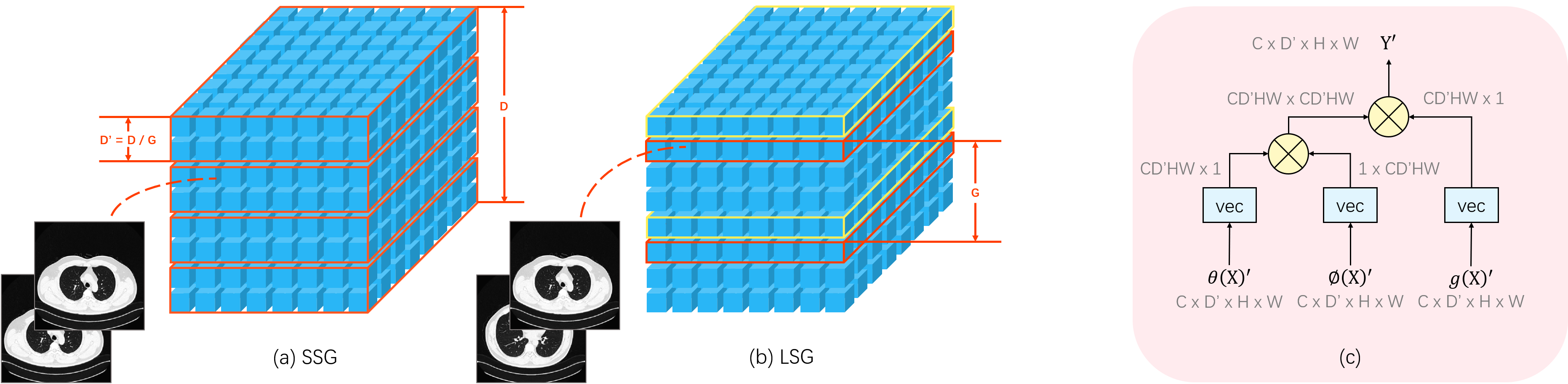}
\caption{Long short slice grouping (LSSG) operation. (a) Short-distance slice grouping (SSG). Embeddings (blue cubes) are grouped by red boxes along the depth dimension. (b) Long-distance slice grouping (LSG). Embeddings with the same color borders belong to the same group. (c) The detailed LSSG operation in one group.} 
\label{LSSG}
\end{figure}

In the thoracic CT images, pulmonary nodules usually show up as isolated spherical shapes, while other tissues like vessels and bronchus appear as the continuous pipe-like structures. Thus, looking at a few successive scans and examining the correlation among them is enough for doctors to differentiate nodules from other tissues. 
Meanwhile, viewing multiple slices also helps with the elimination of continuous pipe-like structures and the localization of nodules. Considering these two aspects, we propose a long short slice grouping (LSSG) operation based on the non-local operation in \cite{non-local18,compact_non-local18},  which is illustrated in Fig. \ref{LSSG}.

We first begin by reviewing the original non-local operation \cite{non-local18}. Suppose $\mathbf{X} \in \mathbb{R}^{C \times D \times H \times W}$ is the input feature map, where $C$, and $D$, $H$, $W$ denote the number of channels, and its depth, height, width. The definition of the non-local operation in \cite{non-local18} is:
\begin{equation}
\mathbf{Y} = f(\theta(\mathbf{X}), \phi(\mathbf{X})) g(\mathbf{X}),
\end{equation}
where $\mathbf{Y} \in \mathbb{R}^{C \times D \times H \times W}$. $\theta(\cdot), \phi(\cdot), g(\cdot) \in \mathbb{R}^{C \times D H W}$ come from $1\times1\times1$ convolutional transformations, and they are defined as:
\begin{equation}
\theta(\mathbf{X}) = \mathbf{W}_{\theta} \mathbf{X}, \phi(\mathbf{X}) = \mathbf{W}_{\phi} \mathbf{X}, g(\mathbf{X}) = \mathbf{W}_{g} \mathbf{X},
\label{transformations}
\end{equation}
where $\mathbf{W}_{\theta}$, $\mathbf{W}_{\phi}$, and $\mathbf{W}_{g}$ are learnable weight matrices. The function $f(\cdot, \cdot)$ implements the  computation between all positions in the feature embeddings. There are several choices for $f$ in \cite{non-local18}. In this paper, we adopt the simple dot-product version:
\begin{equation}
f(\theta(\mathbf{X}), \phi(\mathbf{X})) = \theta(\mathbf{X})^{T} \phi(\mathbf{X}).
\end{equation}
Through this original non-local operation \cite{non-local18}, the long-range dependencies among any locations in the feature map can be captured. Nonetheless, the affinity between any channels is not included, which is demonstrated to be vital for the fine-grained object differentiation \cite{compact_non-local18}. Thus, we merge channel into position to reshape the output of Eq.~(\ref{transformations}), and obtain vectors $vec(\theta(\cdot)), vec(\phi(\cdot)), vec(g(\cdot)) \in \mathbb{R}^{C D H W}$ for the next similarity computation. By doing so, the compact non-local operation can absorb richer long-range dependencies among any elements. Our LSSG operation calculates the response $\mathbf{Y}$ as:

\begin{equation}
\mathbf{Y} = f(vec(\theta(\mathbf{X})), vec(\phi(\mathbf{X}))) vec(g(\mathbf{X})),
\end{equation}
where $vec$ denotes vector reshape operation.
Yet, there is a $CDHW \times CDHW$ pairwise matrix; therefore the computational or memory cost of LSSG is vast. Thus, it is infeasible to directly implement the LSSG operation. 

Some recent works adopt the idea of grouping, e.g. Xception \cite{Xception17}, MobileNet \cite{mobilenets17}, ResNeXt \cite{ResNeXt17}, Group normalization \cite{GN18}, and CrossFormer \cite{crossformer21}. It has been demonstrated that channel grouping or embedding grouping is effective in improving the network performance. This group idea can also be introduced in our LSSG, and considering the properties of pulmonary nodule detection using CT images, we split the LSSG into two parts: short-distance slice grouping (SSG) and long-distance slice grouping (LSG). For SSG, we group the adjacent embeddings along the depth dimension into $G$ groups; each group contains $D^{'}=D/G$ depths. Fig.\ref{LSSG} gives an example where $G=4$. For LSG, the embeddings are sampled along the depth direction with a fixed interval $G$. For instance, in Fig.\ref{LSSG}, all embeddings with a red border belong to a group, and those with a yellow border belong to another. The resulting number of groups is also $G$, and each group contains $D^{'}=D/G$ depths as well. After grouping embeddings, both SSG and LSG employ the LSSG operation within each group:
\begin{equation}
\mathbf{Y^{'}} = f(vec(\theta(\mathbf{X})^{'}), vec(\phi(\mathbf{X})^{'})) vec(g(\mathbf{X})^{'}),
\end{equation}
where $\mathbf{Y^{'}} \in \mathbb{R}^{C \times D' \times H \times W}$, and $vec(\theta(\mathbf{X})^{'}), vec(\phi(\mathbf{X})^{'}), vec(g(\mathbf{X})^{'}) \in \mathbb{R}^{C D^{'} H W}$.
The nodule usually appears in a few successive CT scans, making it unnecessary to utilize all depths for detecting the nodule, whereas depth sparse slices help with the elimination of tissues other than nodules and the localization of nodules. These two situations are nicely covered by two slice groupings of our LSSG, where the similarity between any positions and any channels across slices in long or short distance can be absorbed; thus it facilitates the distinguishment of nodules with various sizes.

The LSSG is then wrapped into the LSSG block, whose definition is:
\begin{equation}
\mathbf{Z} = recover(GN(\mathbf{W}_{z}\mathbf{Y}^{'})) + \mathbf{X},
\end{equation}
where $\mathbf{W}_{z}$ denotes the weight matrx to be learned for a $1\times1\times1$ convolution, and $GN$ is a Group Normalization \cite{GN18}. $recover$ denotes that positions of the slices in all groups are recovered in the depth direction. To make the LSSG block compatible with the existing neural network, the residual connection is adopted.

With regard to the LSSG block configuration, SSG and LSG blocks are added into the pulmonary nodule detection backbone alternately to construct our proposed LSSANet, as shown in Fig. \ref{LSSANet}. It is noteworthy that SANet is a special case of our LSSANet when all the blocks added into the backbone are SSG blocks.

\subsection{Long Short Slice-Aware Network}
\subsubsection{Encoder-Decoder Backbone.} Following prior work \cite{sanet}, we adopt 3D ResNet50 \cite{ResNet16} as the encoder considering its outstanding feature extraction performance. The decoder comprises two $2\times2\times2$ deconvolutional layers; thereby the feature map can be upsampled to proper sizes. Furthermore, two output feature maps from the encoder layers are individually modulated by a $1\times1\times1$ convolutional layer, and then concatenated with the corresponding output of the decoder layers. Our proposed LSSG blocks are integrated into the second and third stages of the ResNet50 encoder (2 to the second and 3 to the third), with SSG and LSG appearing alternately. The U-shaped encoder-decoder architecture \cite{unet15} is as shown in Fig. \ref{LSSANet}.

\subsubsection{3D Region Proposal Network.} In order to generate highly sensitive pulmonary nodule candidates, a $3\times3\times3$ convolutional layer is first applied to the output of the backbone. It is subsequently followed by two parallel $1\times1\times1$ convolutional layers for predicting the classification probability associated with each anchor at each voxel and regressing the 3D bounding box. Each anchor needs to specify six parameters: central point coordinates, width, height, and depth. In this work, five cube anchors with sizes 5, 10, 20, 30, and 50 are chosen. We use the multi-task loss objective in \cite{sanet} for this stage.

\subsubsection{False Positive Reduction.} The nodule candidates generated in the last stage usually contain many false positives. As shown in Fig. \ref{LSSANet}, the feature maps produced by a shallow block in ResNet50 \cite{ResNet16} and the last block of the backbone are cropped using nodule candidates; then the latter feature map is upsampled and concatenated with the former one to obtain the final multi-scale region of interest (RoI). A 3D max pooling is then applied to this RoI. Eventually, two fully connected layers are utilized for generating classification probability and 3D bounding box regression terms. Here we utilize the same loss function as the above 3D region proposal network's. It can not only decrease the number of false positives but also further refine the bounding box regression.

\section{Empirical Study}
In this section, we evaluate our LSSANet's pulmonary nodule detection performance on the large-scale pulmonary nodule dataset PN9 \cite{sanet} using the Free-Response Receiver Operating Characteristic (FROC). Several state-of-the-art approaches are compared, including algorithms based on 2D CNN Faster RCNN \cite{faster_rcnn}, RetinaNet \cite{retinanet}, SSD512 \cite{ssd512}, and algorithms based on 3D CNN Leaky Noisy-OR \cite{leaky_noisy-or}, 3D Faster RCNN \cite{deeplung}, DeepLung \cite{deeplung}, NoduleNet \cite{nodulenet}, I3DR-Net \cite{i3dr-net}, DeepSEED \cite{deepseed}, SANet \cite{sanet}. 

\textbf{Dataset.} We use PN9 \cite{sanet} for evaluating the performance of LSSANet. This dataset is so far the largest and most diverse one for the detection of pulmonary nodules to the best of our knowledge, which comprises 8,796 CT images and 40,436 annotated nodules of 9 different classes. To entail a fair comparison, the 8,796 CT images are split into 6,037 for training, 670 for validation, and 2,089 for testing as in \cite{sanet}. By the way, the PN9 is a preprocessed version, so there are no more data preprocessing steps needed.

\textbf{Implementation Details.} We use the model in \cite{sanet} pretrained on PN9 \cite{sanet} to initialize the weights of our LSSANet. The Stochastic Gradient Descent (SGD) optimizer is applied for training, with the batch size of 16, the learning rate of 0.001, the momentum of 0.9, and the weight decay of $1\times10^{-4}$. The LSSANet is trained for 9 epochs before RCNN gets involved. All the other hyperparameters and experimental settings are the same with \cite{sanet}. Besides, we use a PyTorch implementation of our method. The experiments are conducted on 3 NVIDIA GeForce RTX 3090 GPU with 24GB memory.

\begin{table}
\centering
\caption{Comparison of pulmonary nodule detection performance on PN9 dataset in terms of FROC. Values in each column represent the pulmonary nodule detection sensitivities (unit: \%) under different average numbers of false positives per CT image.}
\label{comparison}
\begin{tabular}{l|c c c c c c c|c} \hline
Method & 0.125 & 0.25 & 0.5 & 1.0 & 2.0 & 4.0 & 8.0 & Avg \\ \hline
\textbf{2D CNN Based Algorithms:} &  &  &  &  &  &  &  & \\
Faster RCNN \cite{faster_rcnn} & 10.79 & 15.78 & 23.22 & 32.88 & 46.57 & 61.94 & 75.52 & 38.10\\
RetinaNet \cite{retinanet} & 8.42 & 13.01 & 20.13 & 29.06 & 40.41 & 52.52 & 65.42 & 32.71\\
SSD512 \cite{ssd512} & 12.26 & 18.78 & 28.00 & 40.32 & 56.89 & 73.18 & 86.48 & 45.13\\ \hline
\textbf{3D CNN Based Algorithms:} &  &  &  &  &  &  &  & \\
Leaky Noisy-OR \cite{leaky_noisy-or} & 28.08 & 36.42 & 46.99 & 56.72 & 66.08 & 73.77 & 81.71 & 55.68 \\
3D Faster RCNN \cite{deeplung} & 27.57 & 36.59 & 46.76 & 58.00 & 70.00 & 80.02 & 88.32 & 58.18 \\
DeepLung \cite{deeplung} & 28.59 & 39.08 & 50.17 & 62.28 & 72.60 & 82.00 & 88.64 & 60.48 \\
NoduleNet \cite{nodulenet} & 27.33 & 38.25 & 49.40 & 61.09 & 73.11 & 83.28 & 89.93 & 60.33 \\
I3DR-Net \cite{i3dr-net} & 23.99 & 34.37 & 46.80 & 60.04 & 72.88 & 83.60 & 89.57 & 58.75 \\
DeepSEED \cite{deepseed} & 29.21 & 40.64 & 51.15 & 62.20 & 73.82 & 83.24 & 89.70 & 61.42 \\
SANet \cite{sanet} & 38.08 & 45.05 & 54.46 & 64.50 & 75.33 & 83.86 & \textbf{89.96} & 64.46 \\ \hline
LSSANet & \textbf{51.59} & \textbf{51.59} & \textbf{58.18} & \textbf{66.88} & \textbf{77.33} & \textbf{85.35} & 89.87 & \textbf{68.69} \\ \hline
\end{tabular}
\end{table}

\textbf{Nodule Detection Performance.} The experimental results are listed in Table \ref{comparison}. As we can see, LSSANet obtains the highest average FROC score over other methods, leading to an improvement of 4.23\% over the second-best SANet, which is actually a special case of our LSSANet. It is noted that our LSSANet surpasses the compared detection approaches to a large extent at all the average numbers of false positives per CT image except for the 8.

\begin{table}
\begin{floatrow}
\capbtabbox{
\centering
\setlength{\tabcolsep}{8mm}{
\begin{tabular}{c|c} \hline
SSG/LSG & FROC \\ \hline
5/0 & 64.46 \\
3/2 & 68.24 \\
2/3 & \textbf{68.69} \\
0/5 & 68.05 \\ \hline
\end{tabular}
}}
{\caption{Effect of implementing with different configurations for the proposed LSSG module (\%). The SSG/LSG denotes the number of SSG vesus LSG.}
\label{ablation}
}
\capbtabbox{
\centering
\setlength{\tabcolsep}{8mm}{
\begin{tabular}{c|c} \hline
Groups & FROC \\ \hline
2 & 68.27 \\
4 & \textbf{68.69} \\
8 & 66.29 \\ \hline
\end{tabular}
}}
{\caption{Effect of using different number of Groups $G$ for the proposed LSSG module (\%).}
\label{ablation_group}
}
\end{floatrow}
\end{table}

\begin{figure}
\RawFloats
\begin{floatrow}
\capbtabbox{
\centering
\setlength{\tabcolsep}{8mm}{
\begin{tabular}{c|c} \hline
Module & FROC \\ \hline
NL-LSSG & 68.30 \\
CNL-LSSG & \textbf{68.69} \\ \hline
\end{tabular}
}}
{\caption{Ablation study for compact non-local operation(\%).}
\label{ablation_module}
}
\figurebox{
\centering
\includegraphics[width=1.8in]{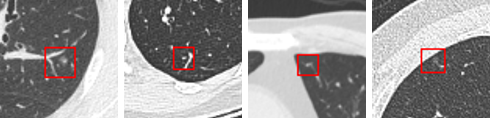}
}
{\caption{Visualization of some pipe-like false positives that SANet detects.}
\label{fp}
}
\end{floatrow}
\end{figure}

\textbf{Different Configurations of LSSG.} Table \ref{ablation} gives the results of our proposed LSSG module using different configurations. We conduct the experiments by making the LSSG blocks in the LSSANet encoder be all SSG blocks, all LSG blocks, or alternate back and forth. The results of LSSANet with 3 LSG and 2 SSG blocks appearing alternately in the encoder are the best, which improves the LSSANet with all SSG blocks and the one with all LSG blocks by 4.23\% and approximately 1\% respectively, in terms of average FROC score. The increase in overall performance could be attributed to the ability to focus on the small-scale features of nodules existing in consecutive slices, and extract the global features from the sparse slices.

We also analyze the sensitivity of our proposed LSSG w.r.t. different numbers of groups $G$, as listed in Table. \ref{ablation_group}. The number of groups 4's FROC score is the best, which improves the setting $G=8$ by over 2\%.

\textbf{Effectiveness of CNL.} We add an ablation study to compare the network equipped with the compact non-local operation (CNL-LSSG) with the one utilizing the original non-local operation (NL-LSSG). From Table. \ref{ablation_module}, we can see that the compact non-local operation is beneficial for the pulmonary nodule detection.

\textbf{Visualization Analysis.} Here we provide visualization of some pipe-like structures in Fig. \ref{fp}, which have high probabilities to be detected as false positives by the original SANet. In contrast, our LSSANet is able to discriminate the nodules from continuous pipe-like structures (such as blood vessels) beyond a few neighboring slices.

\section{Conclusion}
In this paper, we develop a long short slice-aware network (LSSANet) for detecting the pulmonary nodules. A novel long short slice grouping (LSSG) module is proposed to explore explicit correlations among any positions and any channels of long-distance or short-distance slice groups. Experimental results on the large-scale dataset PN9 \cite{sanet} demonstrate that LSSANet can achieve the state-of-the-art performance in pulmonary nodule detection, and both the long-distance slice grouping (LSG) and short-distance slice grouping (SSG) are critical in improving the performance. Besides, our LSSG is flexible and can be plugged into many nodule detection models. In the future, we intend to incorporate the proposed LSSG in more nodule detection frameworks to improve their performance.

\subsubsection{Acknowledgements.} This work was partially supported by National Natural Science Foundation of China under Grant 62141112, and the Special Fund of Hubei Luojia Laboratory under Grant 220100014. This work was also supported in part by National Science Foundation of China (82071990, 61976238).

\bibliographystyle{splncs04}
\bibliography{paper2658}

\end{document}